\documentclass[a4paper,11pt]{article}
\usepackage{booktabs}
\usepackage{bm}
\usepackage{float}
\usepackage{latexsym}
\usepackage{dcolumn}
\usepackage{amsfonts,amssymb}
\usepackage{graphicx}
\usepackage{epsfig}
\usepackage{psfrag}
\usepackage{nccmath}
\usepackage{moresize}
\usepackage{enumerate}
\usepackage[hang,flushmargin]{footmisc}
\usepackage[titletoc,toc]{appendix}
\usepackage{lipsum}
\usepackage{subcaption}
\usepackage{hyperref}
\usepackage{rotating}
\usepackage{multirow}
\usepackage{multicol}
\usepackage{xfrac}
\usepackage{mathtools}
\usepackage{nccmath}
\usepackage{cite}
\usepackage{placeins}
\usepackage{tikz}
\usetikzlibrary{positioning}
\usepackage{textalpha}
\graphicspath{{figures/}} 
\usepackage[font=small,labelfont=bf]{caption}
\usepackage{wrapfig}
\usepackage{authblk}
\usepackage{cite}
\usepackage[utf8]{inputenc}
\usepackage[T1]{fontenc}
\usepackage{enumitem}
\usepackage{blindtext}

\title{\bf Phantom dark energy as a natural selection of evolutionary processes\\ $\hat{\rm a}$  {\it  la} {\it genetic algorithm} and cosmological tensions}   

\author{Mayukh R. Gangopadhyay\thanks{mayukh\_ccsp@sgtuniversity.org}}
\author[2]{\large M.~Sami${}^{1,2,3}$ 
\thanks{ sami\_ccsp@sgtuniversity.org,  samijamia@gmail.com}}
\author[3]{\large Mohit K. Sharma${}^{1}$
\thanks{mr.mohit254@gmail.com }}
\affil{\small ${}^{1}$Centre For Cosmology and Science Popularization (CCSP),
        SGT University, Gurugram,  Haryana- 122505, India,}
\affil{\small ${}^{2}$Eurasian International Centre for Theoretical Physics, Astana, Kazakhstan.}
\affil{\small ${}^{3}$Chinese Academy of Sciences,52 Sanlihe Rd, Xicheng District, Beijing.}

\date{}

\begin{document}

\maketitle

\begin{abstract}
    We study the late-time cosmological tensions using the low-redshift 
    background and redshift-space distortion data by employing a machine 
    learning (ML) technique. By comparing the generated observables with the 
    standard cosmological scenario, our findings indicate support for 
    the phantom nature of dark energy, which ultimately leads to a reduction 
    in the existing tensions. The model-independent approach also enables 
    us to examine the combined background and perturbative history, where 
    tensions are reduced. Moreover, from a statistical 
    perspective, we have shown that our results exhibit a better fit to the 
    data when compared to the $\Lambda$CDM model.
\end{abstract}

\section{Introduction}

One of the major challenges in modern cosmology such as the discrepancies 
between high-redshift observations of the Cosmic Microwave Background (CMB) \cite{AT-book,planck18}, 
and low-redshift surveys such as galaxy clustering and weak gravitational lensing \cite{riess,holicow,kids450,kids1000}, prompt one to pose a question: {\it whether the 
Cosmological Constant ($\Lambda$) can be considered a plausible candidate for 
Dark Energy (DE)?} As supported by most of the cosmological observations, the 
$\Lambda$CDM model (where CDM refers to the Cold Dark Matter) has
recently been subjected to intense scrutiny, particularly with respect to the 
identification of a high expansion rate and less matter-density clustering in 
the low-redshift observations. For an instance, low-redshift observations 
such as Supernovae $H_0$, for the Equation of State of DE (SH0ES) 
\cite{riess} and Kilo Degree Survey (KiDS) \cite{kids450,kids1000} 
have challenged Planck-18 estimates for the Hubble constant ($H_0$) 
and the matter density clustering ($\sigma_8^{(0)}$) by revealing 
discrepancies of about $5 \sigma$ and $3 \sigma$, respectively 
\cite{val-rev,elen,running-H0sig8,dain-hTension,dain-hTension2,nunes,cpsingh,huterer}. 
These inconsistencies are not limited to the $\Lambda$CDM model but 
also extend to weakly dynamical Dark Energy (DE) models that mimics it. 
As a substantial range of models unable to address these disparities, 
there are two potential explanations for this: either there exists a 
systematic error in the data or the $\Lambda$CDM model is not a 
suitable one.

In the literature, many alternative approaches, including those based on 
Modified Gravity (MG) theories have investigated the issue by
considering the possibility that the $\Lambda$CDM model itself may be 
responsible for these inconsistencies. These approaches includes 
interactions between dark energy and dark matter 
\cite{amendola,sujikawa,pandey}, modifications of gravity at early 
\cite{trodden,trodden2,sunny1} or late times 
\cite{generic,generic2,mont-htension}, 
distinctive perspective on the dynamic vacuum energy 
(DVE) concerning the dynamical dark energy
\cite{sola-1,sola-2,sola-3,sola-4,sola-5}
and so on \cite{silk,abdalla,elenora,sigma8drag,pogosian,perivola,supratik, neserris2, neserris3,colgain,souza-neutrino,neutrino-tensions,chan-hTension,daniel-s8tension,dain-hTension,clark-H0S8Tension,AE-tension,KP-tens,tsallis}. Some of these approaches also consider phenomenologically 
constructed DE models. However, despite the attempts to explain the 
low-redshift data, these models often include inherent biases and 
assumptions. Therefore it can also introduce biases in the estimation of 
cosmological observables, which may lead to model-specific results 
rather than the one which can be largely applicable. Due to this 
potential lack of concordance among several models, the investigation 
of the tensions necessitates the consideration of a model-independent 
approach.


In this paper, we adopt a novel model-independent technique that only 
relies on the data such as the cosmological background and linear 
perturbative level to study the evolution of the universe.
In particular, {\it we use a population-based metaheuristic optimization 
algorithm that is inspired by the process of natural selection, 
wherein a fitness function evaluates the 
fitness (such as the goodness of fit) of individuals (potential solutions) 
at each step \cite{GA1,GA2,GA3,GA4,GA5}. This approach makes use of multiple potential solutions. 
 It ensures that the population (number of solutions) maintains diversity 
and prevents the optimal solution from becoming trapped in local minima.
Once the individuals from the population are selected to reproduce,
offspring of the next generation, they may merge together or get themselves 
modified to enhance fitness. This process continues in an effort to 
emulate the process of natural selection. }

To start, we need an initial group of randomly 
created mathematical functions. Each function in 
the population is evaluated using a fitness 
function that measures how well the function fits 
the given data points. Our approach involves 
utilizing the $\chi^2$ statistic as the fitness 
function to assess the 
``success'' of reproducing next generation 
solutions. After evaluating the fitness of each 
function in the current population, the ones with 
higher fitness are selected to become parents for 
the next generation. Once this step is completed, 
pairs of different functions take part in 
crossover to create new functions and the old 
nodes of the tree gets replaced by the new ones. 
Mutation is also an essential step in this process 
because it changes the functions and checks if 
these changes improve their fitness. The whole 
process continues until it either reaches the 
maximum number of generations or when the 
functions achieve their highest fitness level.

The degree of effectiveness for each solution, which just depends 
on a single independent variable like redshift, is based on its 
ability to align with the observational data. The population which 
minimizes the $\chi^2$ is considered to have the highest fitness. Therefore, 
our stochastic process aims to proceed in the direction of minimization.
It is worth noting that the accuracy of the final optimal solution is largely 
unaffected by changes in the initial population. Regardless of the initial 
conditions, the optimization process will generally lead to the same optimal 
solution, unless any singularities are encountered along the way. The main 
advantage of using this method is that it can automatically discover most 
relevant complex features from the data which may be beyond the capability of 
standard parametric methods.

Our main objective in this paper is to determine if the optimal solution 
deviates from the $\Lambda$CDM model, and if it does, whether it also 
alleviates the cosmological tension(s). To apply the aforementioned 
technique to simulate the process of natural selection for the desired 
observables we intend to use the cosmological background and 
redshift-space distortion (RSD) data. Since in our approach there are no 
parameters, so extracting the required DE information from the optimal 
functional form is highly non-trivial. To determine the cosmological 
parameters, we choose a cosmological model that encompasses a wide range 
of DE models and attempt to fit this model to the optimal solution. 
An advantage of this approach is that when the optimal solution is already 
identified, the goodness of fit of the optimal solution is typically 
significantly better than that of the parametric methods. This is due to the 
pre-defined functional form in the latter. Once the optimal solution 
is obtained, a chosen cosmological model can be mapped with it which can  
essentially provide those fitted parametric values of the model that 
align with the optimal solution.

The outline of this paper is as follows: Firstly, we study the 
background expansion rate and then obtain the background optimal solution 
for the Hubble parameter, we then obtain the corresponding DE equation 
of state and matter density parameter. Secondly, we implement the algorithm 
on the RSD dataset, and obtain the corresponding cosmological parameters. 
Based on the estimates, we then obtain the bounds on the $S_8$ parameter. 
Finally, by using the algorithm assisted optimal solutions for the 
expansion rate and growth of matter density perturbations we obtain a 
unified trajectory between them which can be treated as a optimal one 
in which the tension is reduced or absent.

\section{Cosmological background evolution}

In this section we will analyze the cosmological background data by using the genetic algorithm (GA) approach. The main objective to consider the GA approach is to remove any biases or the assumptions associated with a chosen cosmological model. This allows one to look for hidden information in the data without encounter the constraints of a cosmological framework. For an example, in the $\Lambda$CDM model, the present-day values of the Hubble parameter or the Hubble constant $H_0$ from the local distance ladder and the cosmic microwave background (CMB) measurements differ by an almost $4\sigma$ level. This allows one to reconsider the choice of the $\Lambda$CDM model. Hence, 
without resorting to a particular model, one can have a better understanding of the underlying data. Our main aim is to identify the patterns in the data using a population-based 
algorithm that can reveal features not easily noticeable in standard 
cosmological setup. For this reason, we will utilize two different data 
sets for the background analysis to figure out what does the data actually 
infer.

\paragraph{Data set-1:}
In order to execute the aforementioned algorithm, 
in this case we use two datasets: ($i$) Observational Hubble data (OHD) from different redshifts in the range
$0.07 < z < 1.965$. In particular, we consider a compilation of $31 \, H(z)$ measurements obtained
from the cosmic chronometric (CC) method (enlisted in \cite{generic}). 
The main reason to use 
the CC dataset is that it provides direct information about the Hubble parameter at different 
times (redshifts). This is different from other methods that only measure quantities 
like luminosity distances etc. without directly studying $H(z)$.
($ii$) For the SN1a dataset, we make use of the Hubble rate denoted as $E(z):= H(z)/H_0$, which 
consists of six data points within the range of redshift $z$ from 0.07 to 1.5. These 
six data points effectively contain the information from a larger set of 1048 data points 
from the Pantheon catalog, as well as 15 data points from the CANDELS and CLASH Multi-Cycle 
Treasury (MCT) programs obtained by the Hubble Space Telescope (HST). Additionally, based 
on the arguments presented in the reference \cite{amendola-GP}, the data point at $z = 1.5$ 
has been excluded from our analysis. 
The execution of the algorithm is done by incorporating the Likelihood function, which is given as
\begin{equation}\label{likelihood}
    L(\chi^2) \propto e^{-\chi^2/2} \, \quad 
    \mbox{such that} \quad 
    \chi^2 = \begin{cases}
    \sum_i \left( \frac{H_{obs} - H_{alg}}{\sigma_i} \right)^2  \quad \mbox{For OHD} \\
    \sum_{i,j}\left[E_i - (h_{alg})_i \right] \cdot c_{ij}^{-1} \cdot \left[E_i - (h_{alg})_i \right] \quad \mbox{For SN1a} \,.
    \end{cases}
\end{equation}
where $H_{obs}$ and $H_{alg}$ denotes the observed and 
algorithm fitted Hubble parameter values, respectively, $c_{ij}$ is the covariance matrix, and 
$h_{alg}$ is the algorithm assisted reduced Hubble parameter. 

\paragraph{Data set-2:} This compilation comprises three datasets: ($i$) 31 
measurements of CC, as previously mentioned. ($ii$) SN1a dataset, for 
which we utilize the latest and most comprehensive Pantheon+ dataset, which includes apparent magnitudes calculated from $1701$ light curves representing $1550$ SN1a events across a redshift range of $z$ spanning from $0.001$ to $2.26$, obtained from 18 different surveys. This dataset represents a substantial improvement compared to the initial Pantheon sample of $1048$ SN1a events \cite{pantheon}, particularly at lower redshift values. 
The theoretical formula for the apparent 
magnitude $m_{B}$, which is related to the Hubble independent 
luminosity distance $D_L$ i.e.
\begin{equation} 
D_L(z) = H_0  d_L(z) \, , \quad \mbox{where} \quad
 d_L(z)= (1+z) \int_0^z \frac{d\tilde{z}}{H(\tilde{z})}  \, ,
\end{equation}
can be expressed as
\begin{equation}
m_{B}(z) = M + 5 \log_{10}[D_L(z)] + 5 \log_{10}\left( \frac{H_0^{-1}}{1 \text{Mpc}}\right) +  25 \, ,
\end{equation}
where $M$ is the absolute magnitude of SN.
The $\chi^2$ for the SN1a is thus can be written as:
\begin{equation}
    \chi^2_{SN} := \Delta m_B \cdot C^{-1}_{SN} \cdot \Delta m_B \,,
\end{equation}
where $C^{-1}_{SN}$ is the inverse of the SN1a covariance matrix. 
($iii$) The BAO dataset enlisted in \cite{ratra}, includes 
measurements of various cosmological parameters such as the Hubble 
distance $D_H(z)$, transverse comoving distance $D_M(z)$, and 
volume-averaged distance $D_V(z)$. These measurements encompass a 
redshift range that ranges from $0.38$ to $2.334$. For this dataset we will 
adopt the parameter $\Omega_b = 0.02242$ \cite{planck18} and utilize the 
sound horizon value at the drag epoch $r_d = 147.78\,$Mpc at $z=1059$ \cite{mifsud}. The corresponding $\chi^2$ is given as:
\begin{equation}
    \chi^2_{BAO} = \Delta A \cdot C^{-1}_{BAO} \cdot \Delta A
\end{equation}
where $A$ represents the observed quantity. The total $\chi^2$ is thus 
expressed as:
\begin{equation}
    \chi^2_{alg} = \chi^2_{OHD} + \chi^2_{SN} + \chi^2_{BAO} \,.
\end{equation}

As one can see that $H_{alg}$ is not defined yet, this is because through the 
evolutionary process we will try to find its best-fit functional form without 
assuming any cosmological restrictions. In order to obtain the desired form 
of $H_{alg}$, we first consider a set of some
individuals in the form of mathematical functions such as polynomials, exponentials,
etc., which goes through a process of merging and modification. A population of $N$ individuals undergoes combinations after each iteration, and their fitness or likelihood is calculated. Here we note that among the population,
individuals with higher fitness are then again considered to generate the new
combinations for the next generation, but at the same time, the individuals having
lower fitness levels are not excluded from this process. As a result, the algorithm has a tendency to continue searching in the continuous search space in such a way that minor modifications can significantly enhance an individual's fitness. Due to this very reason, one can assure that the final solution does not prematurely converge and approaches the global optima rather than the local one.

For the dataset-1, the best-fit solutions $H_{alg}(z)$ with minimum $\chi^2$ value (corresponds to the maximum fitness) is obtained as 
\footnote{Here we mention that while the best-fit $\chi^2$ for each run may vary slightly from other runs or require more generations to converge, it consistently yields an almost indistinguishable cosmological evolutionary scenario.}:

{\underline{For OHD}}:
\begin{equation} \label{H_alg}
    H_{alg}(z) = 72.76 + 68.32 z^2 - 13.03 z^4 + 
    0.001 z^{15} \,\,[km/s/Mpc]\,, \quad \mbox{with} \quad \chi^2 = 12.58 \,.
\end{equation}
{\underline{For OHD + SN1a}}:
\begin{equation} \label{H_alg_SN1a}
    H_{alg}(z) = 70.307 \exp(0.850 z) + 0.769 z^8-5 z^6-35.798 z \,\,[km/s/Mpc]\,, \quad \mbox{with} \quad \chi^2 = 19.72 \,.
\end{equation}
Let us here note that for the $\Lambda$CDM, the $\chi^2$ 
value turns around $14.5$ for the OHD and $21.19$ for OHD+SN1a, 
therefore our result represents 
a significant improvement in the  fit by about $13\%$ and $7\%$ for 
OHD and OHD+SN1a, respectively 
\footnote{It is important to mention that our algorithm's 
non-parametric nature prevents us from utilizing information criteria 
like AIC and BIC. These methods impose penalties based on the number 
of parameters, which is not applicable in our case since our approach 
does not involve such parameters. Therefore, applying AIC or BIC directly 
to non-parametric methods is not straightforward because these criteria 
lack a fixed number of parameters.}.
Here, let us emphasize in order to try to decrease 
variance or over-fitting as well as the likelihood of 
being trapped in the local minima, we have started each 
run with a significant large population $\sim \mathcal{O}(10^4)$.
Also, from Eq.\,(\ref{H_alg}) the present-day ($z=0$) 
best-fit value for the Hubble parameter ($H_{alg}$) is 
determined to be $72.76$ km/s/Mpc. Since, this best-fit 
value fits better with the data as compared to the 
$\Lambda$CDM model, it suggests a strong preference for a 
gravitational modification in the late universe over the 
$\Lambda$CDM and similar cosmological frameworks.

\begin{figure}
\begin{subfigure}{0.495\linewidth} \centering 
   \includegraphics[height=6.7cm,width=8cm]{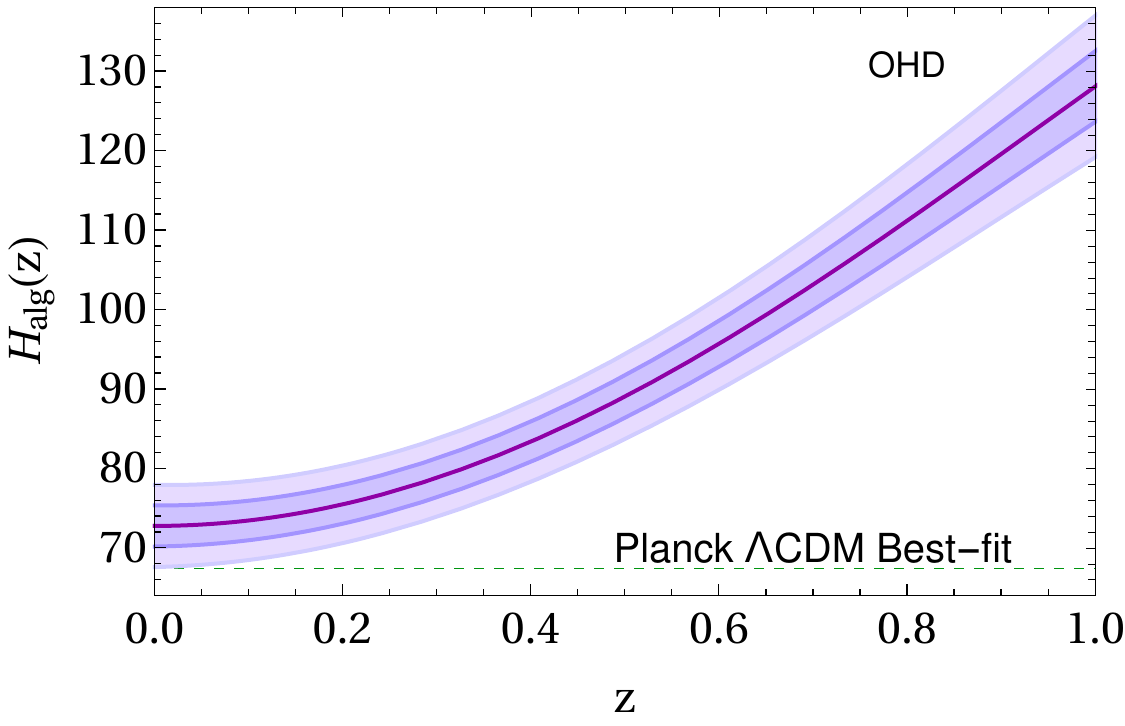}
   \caption{} \label{ga_Hubble}
\end{subfigure}
\begin{subfigure}{0.495\linewidth} \centering
    \includegraphics[height=6.7cm,width=8cm]{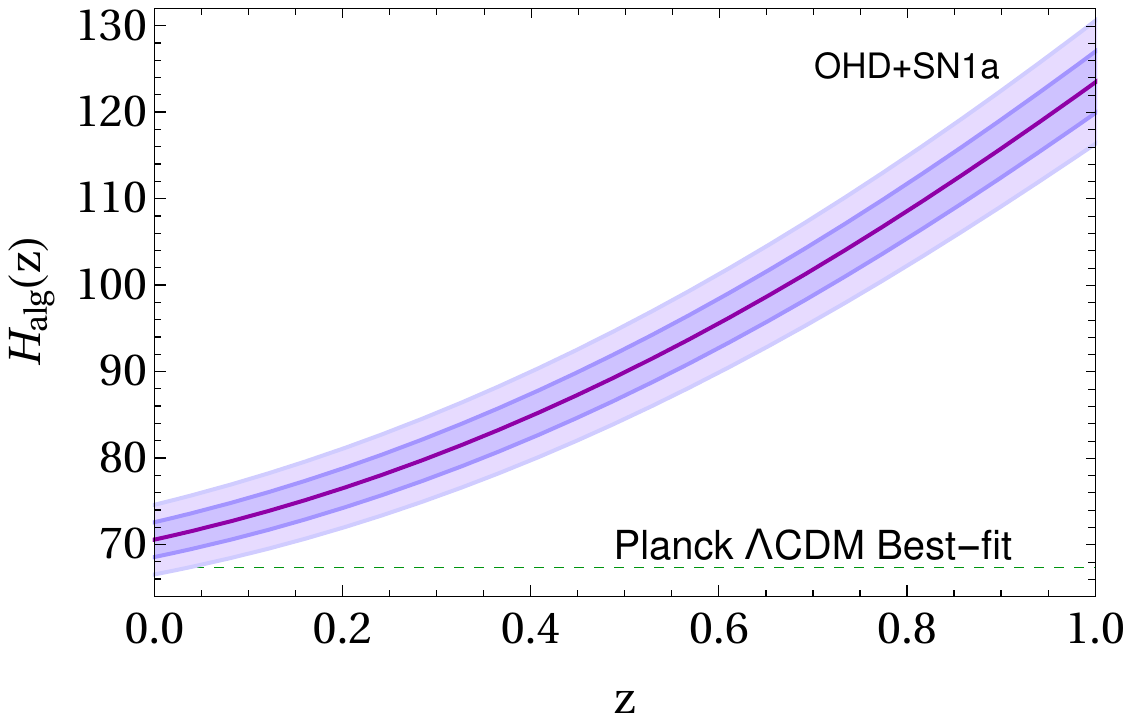}
    \caption{} \label{Hubble-hist}
\end{subfigure}
    \caption{\small For Data set-1, figures (a) and (b) show the evolutionary profiles of $H_{alg}(z)$ with $z \in [0,1]$ upto $2\sigma$ level for OHD and its combination with the SN1a dataset. The solid line represents the best-fit curve. It clearly shows a notable difference in the Hubble expansion rate when compared to the $\Lambda$CDM model, wherein the Hubble rate tends to be comparatively smaller.}
    \label{fig:GA_Hubble}
\end{figure}

In order to obtain the confidence limits for the above 
non-parametric best-fit (\ref{H_alg}) we resort to the 
bootstrap technique for error estimation. In particular, 
it generates multiple bootstrap samples by randomly 
sampling with replacement from the given dataset, and by 
using them we get the standard deviations or confidence 
intervals for our observable. The obtained $2\sigma$ profiles of 
$H_{alg}(z)$ with best-fit values are shown in 
fig.\,(\ref{fig:GA_Hubble}).
While the Hubble parameter in fig.\,(\ref{ga_Hubble}) 
does follow the expected trend 
of decreasing with $z$ at higher redshifts, there is a 
noticeable deviation around $z \simeq 0.2$ where 
the decrement of $H_{alg}(z)$ tends to decrease. This 
contradicts the prediction of various DE models, which 
suggest that the Hubble parameter will continue to 
decrease till it becomes almost constant in the far future 
(which corresponds to the de-Sitter universe). On the other hand, 
in fig.\,(\ref{Hubble-hist}) we show the $H_{alg}(z)$ profile 
for OHD+SN1a dataset. Here, we observe that the error profile becomes 
narrower in the latter case, even then it exhibits a tendency 
towards larger values of $H_0$.
The results of fig.\,(\ref{fig:GA_Hubble}) signals towards 
the fact that if a particular constituent 
of the universe, which may be attributed to DE or a result 
of modified gravity, is intrinsically responsible for the 
enhancement in $H(z)$ through a positive time-derivative, 
it could necessarily exhibit a phantom-like behavior 
\cite{gum}. 

\begin{figure}
\centering
    \includegraphics{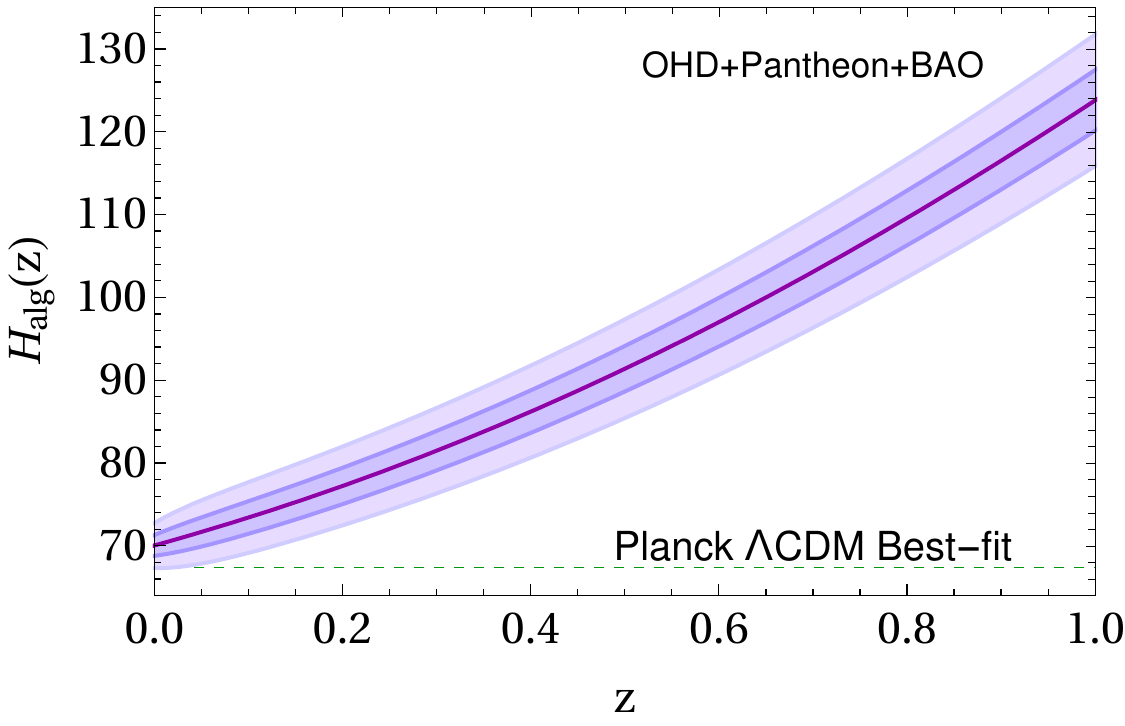}
    \caption{For Data set-2, the figure show the evolutionary profiles of $H_{alg}(z)$ with $z \in [0,1]$ upto $2\sigma$ level.}
    \label{fig:H_comb}
\end{figure}

For the dataset-2, we again follow the same procedure to obtain the 
best-fit functional form of $H_{alg}(z)$ followed by the cosmological 
parametric values. In this case, we obtain the following:
\begin{equation} \label{H_comb}
    H_{alg}(z) = 70.08 \left(1+z \left(0.6715+0.22 z+0.005 z^2-0.029 z^3+0.01 z^4-0.0013 z^6\right)^2\right) [km/s/Mpc]\,,
\end{equation}
from which one finds that the Hubble constant is $70.08\,$km/s/Mpc. 
Whereas, for the $\Lambda$CDM, we have obtained the best-fit value 
of the Hubble constant approximately $68.9\,$km/s/Mpc. This 
represents a substantial improvement of $\Delta H(0) = 1.18$ km/s/Mpc 
in the Hubble constant 
when compared to the $\Lambda$CDM model, thus highlighting its 
significance in addressing the Hubble tension problem. Furthermore, 
there is an almost $\Delta \chi^2 = \chi^2_{\Lambda CDM} - \chi^2 _{alg} \simeq 2$ improvement compared to 
the $\Lambda$CDM model, making our results again more favorable when 
using the combined dataset. It is also worth emphasizing that 
due to the chosen $\Omega_b$ and $r_d$ that corresponds to the 
best-fit result of Planck $\Lambda$CDM, there is some level of 
influence of the $\Lambda$CDM model in the above obtained form of the $H_{alg}$. 
Nevertheless, the obtained profile is depicted in fig.\,(\ref{fig:H_comb}), where it is evident that the Planck $\Lambda$CDM best-fit 
falls within a $2\sigma$ confidence interval. 
Notably, the discrepancy with the SH0ES estimate $73.04 \pm 1.04\,$km/s/Mpc has been reduced to within a $2 \sigma$ level 
when considering the CC with full Pantheon+ and BAO dataset. 

\begin{figure}
    \centering
    \includegraphics[height=7cm,width=14cm]{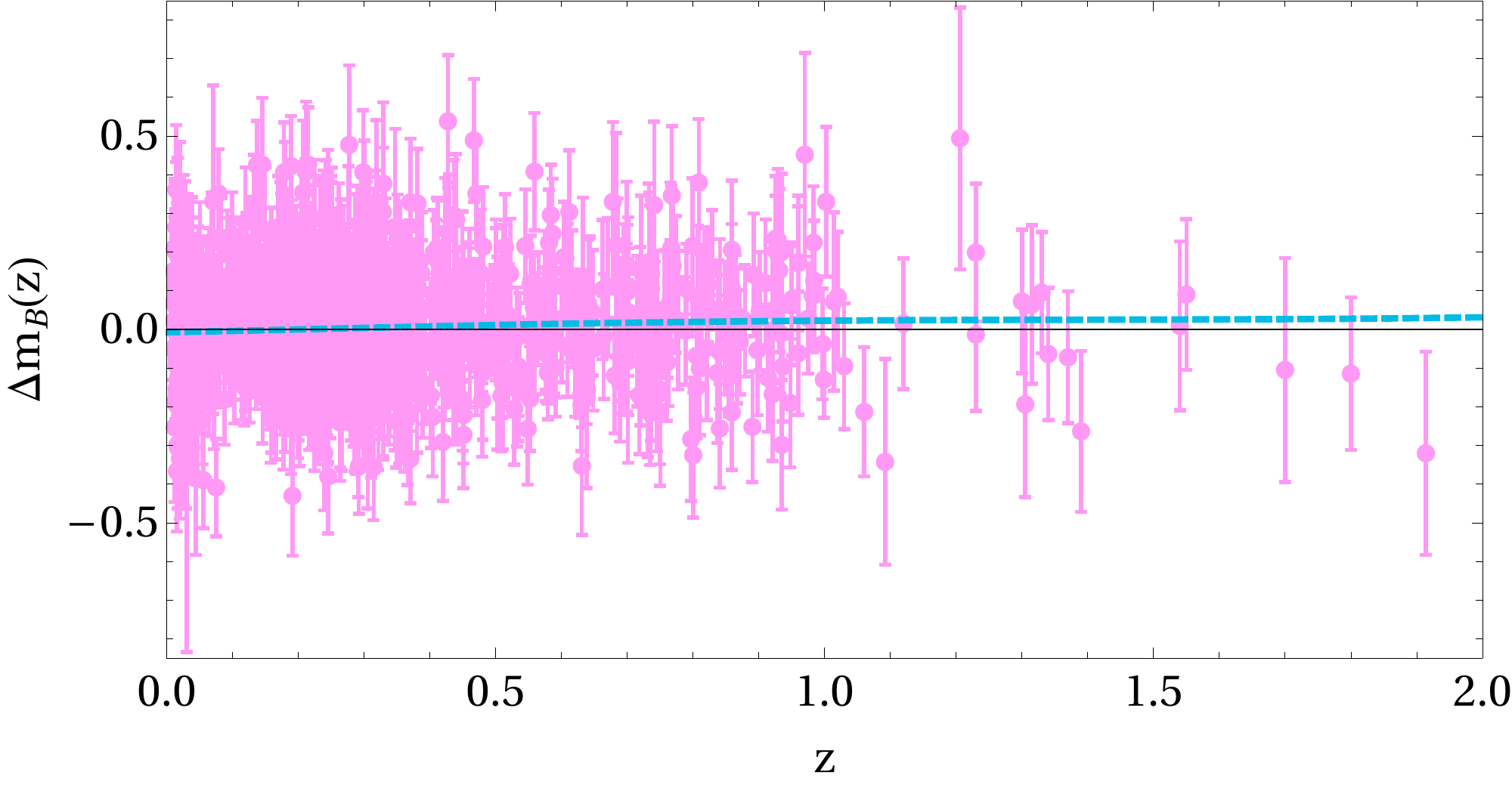}
    \caption{This figure represents the evolution of 
    $\Delta m_B(z)$ with $z \in[0,2]$. The errorbars 
    correspond to the Pantheon+ dataset, and the blue 
    dashed line is the difference between $\Lambda$CDM 
    and our best-fit results obtained from the dataset-2. }
    \label{fig:delta_mb}
\end{figure}

In fig.\,(\ref{fig:delta_mb}) we show the observed $\Delta m_B$ 
for Pantheon+ dataset with upto $z=2$. The blue-dashed line 
represents the difference between the apparent magnitudes of 
for (\ref{H_comb}) and $\Lambda$CDM model. One can see that 
$\Delta m_B$ changes sign around $z=2.5$, preferring 
slightly small $m_B$ near the current epoch than the concordance 
model. This is mainly attributed to the fact that $H_{alg}>H_{\Lambda CDM}$ as can be seen from 
fig.\,(\ref{fig:H_comb}).

To quantitatively measure the effective contribution and dynamical nature of 
dark energy (DE), it becomes essential to employ a standard Hubble parameter 
form. However, our approach doesn't involve any free parameters, which prompts 
the need to select a specific cosmological framework for estimating the 
cosmological parameters which corresponds to the above fit. After obtaining the 
$H(z)$ profiles from two datasets, our next step involves deriving the 
corresponding cosmological parameters out of it.

\subsection{Cosmological background Parameter estimations} \label{background}

To ensure unbiased estimates of the parameters, unaffected by the 
choice of a specific observable form, we compare the evolution of 
$H_{alg}(z)$ with the standard framework of the flat-$w$CDM expression 
of the Hubble parameter, denoted as $\mathcal{H}(z)$. 
The comparison between $\mathcal{H}(z)$ and $H_{alg}(z)$ involves the evaluation of parametric and non-parametric forms of the Hubble parameter to deduce the parametric values. It is important to note that direct inference of cosmological parameters like $w_{DE}$ is not feasible from 
the $H_{alg}$ alone, unless one has the prior knowledge of 
$\Omega_m^{(0)}$ or $\Omega_{DE}^{(0)}$ values \cite{AT-book}.
For a 
fairly general setup 
\footnote{Here we are restricting ourselves for 
a class of theories which does not take into account the features of dynamical 
vacuum energy \cite{sola-1, sola-2, sola-3, sola-4, sola-5}. In these cases, the EoS exhibits quintessence or phantom behavior, through by contributions 
from bosons and fermions in the loop calculation.}, 
the Hubble parameter can be written as \cite{periv2}:
\begin{equation} \label{H_DE}
    \mathcal{H}(z) = H_0 \sqrt{\Omega_{m}^{(0)} (1+z)^3 + (1 - \Omega_{m}^{(0)})\, (1+z)^{1+w_{_{DE}}} } \,, 
\end{equation}
where $H_0$ denotes the Hubble constant, and 
$\Omega_{m}^{(0)}$ represents the the current density 
parameter. Assuming a flat universe with pressure-less 
dust, the total equation of state parameter $w_{_T}$ of 
the system is related to  $w_{_{DE}}$ as 
$w_{_T}=\Omega_{_{DE}} w_{_{DE}}$. 

\begin{figure}[t]
\centering
\begin{subfigure}{0.855\linewidth} \centering 
   \includegraphics[height=3.6cm,width=12cm]{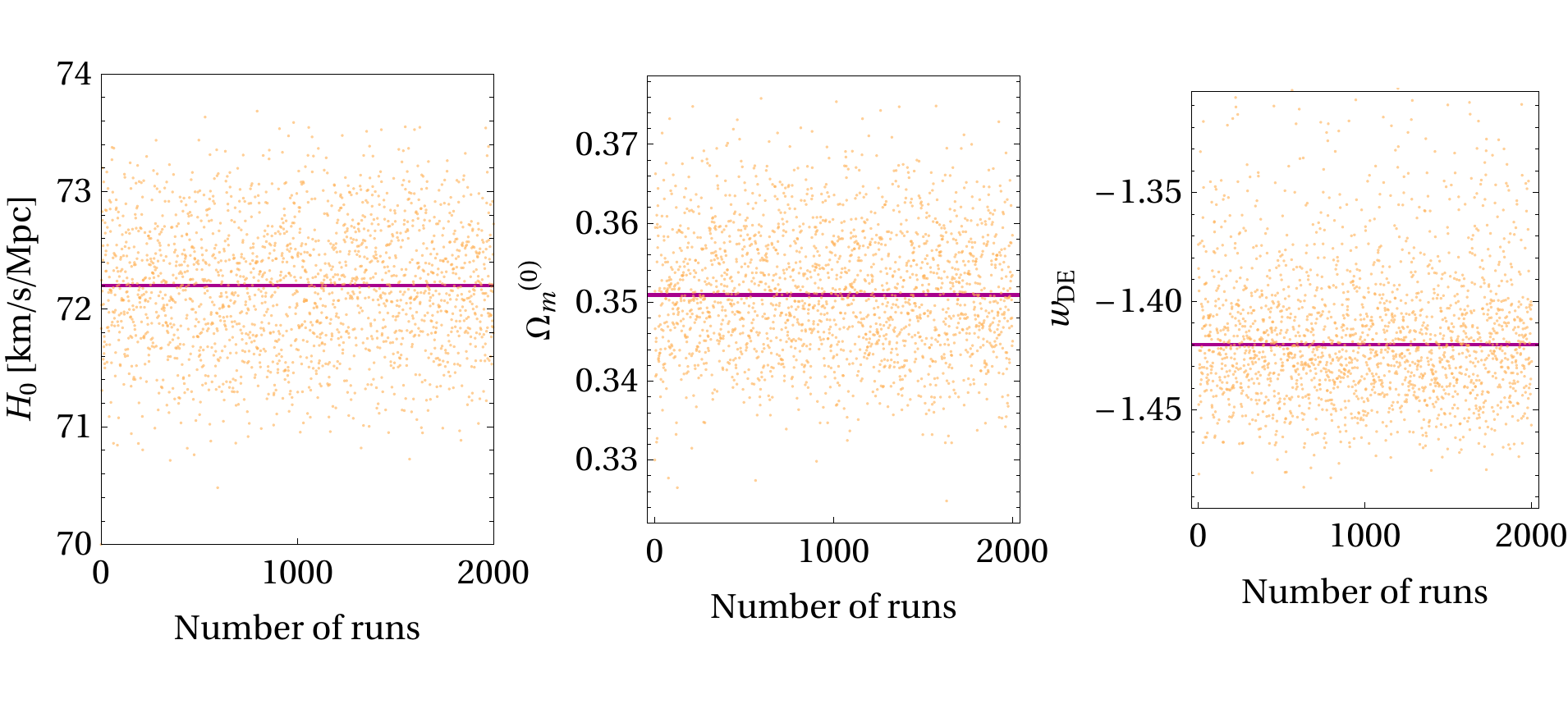}
   \caption{With $H(z)$ data.}
\end{subfigure}
\begin{subfigure}{0.855\linewidth} \centering
    \includegraphics[height=3.6cm,width=12cm]{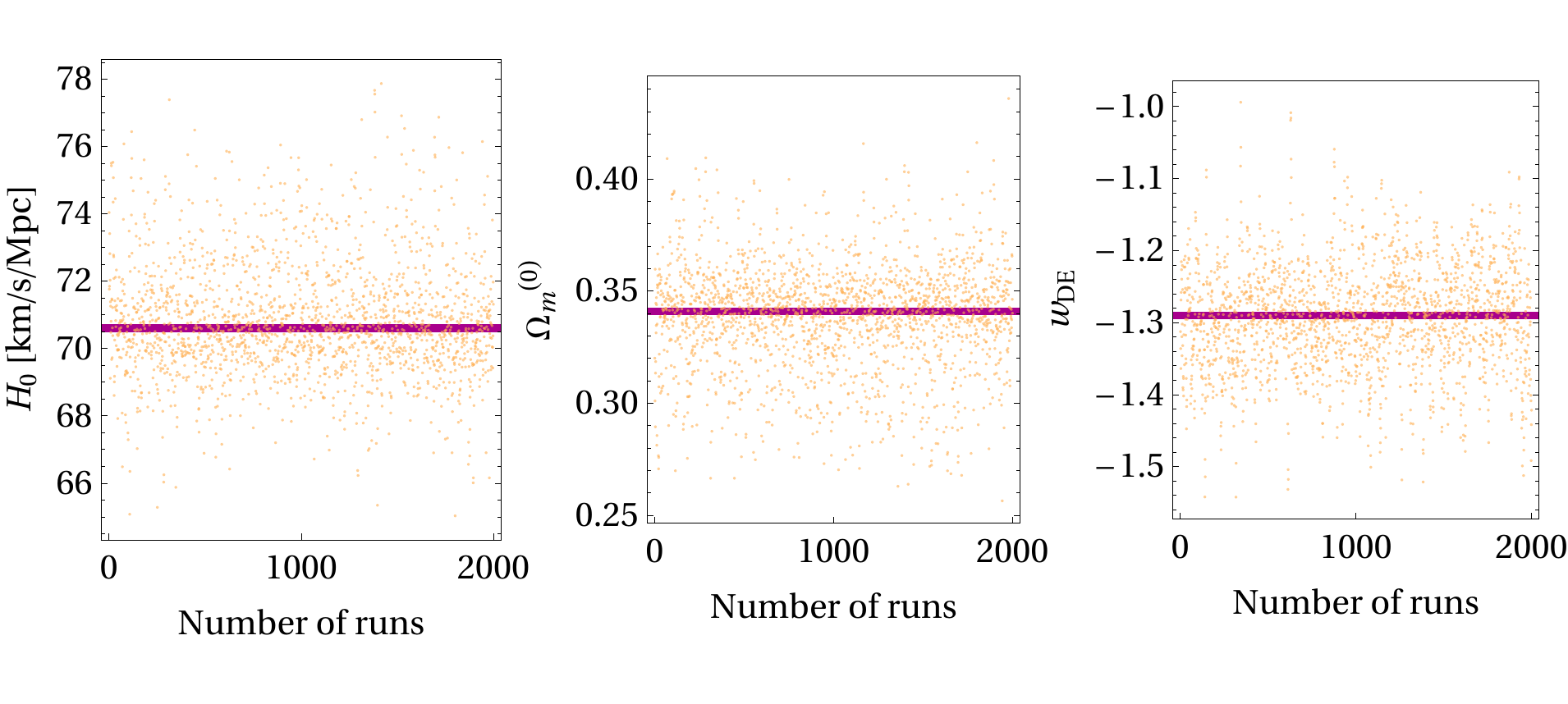}
    \caption{With $H(z)$+SN1a data.} 
\end{subfigure}
\begin{subfigure}{0.855\linewidth} \centering
    \includegraphics[height=3.6cm,width=12cm]{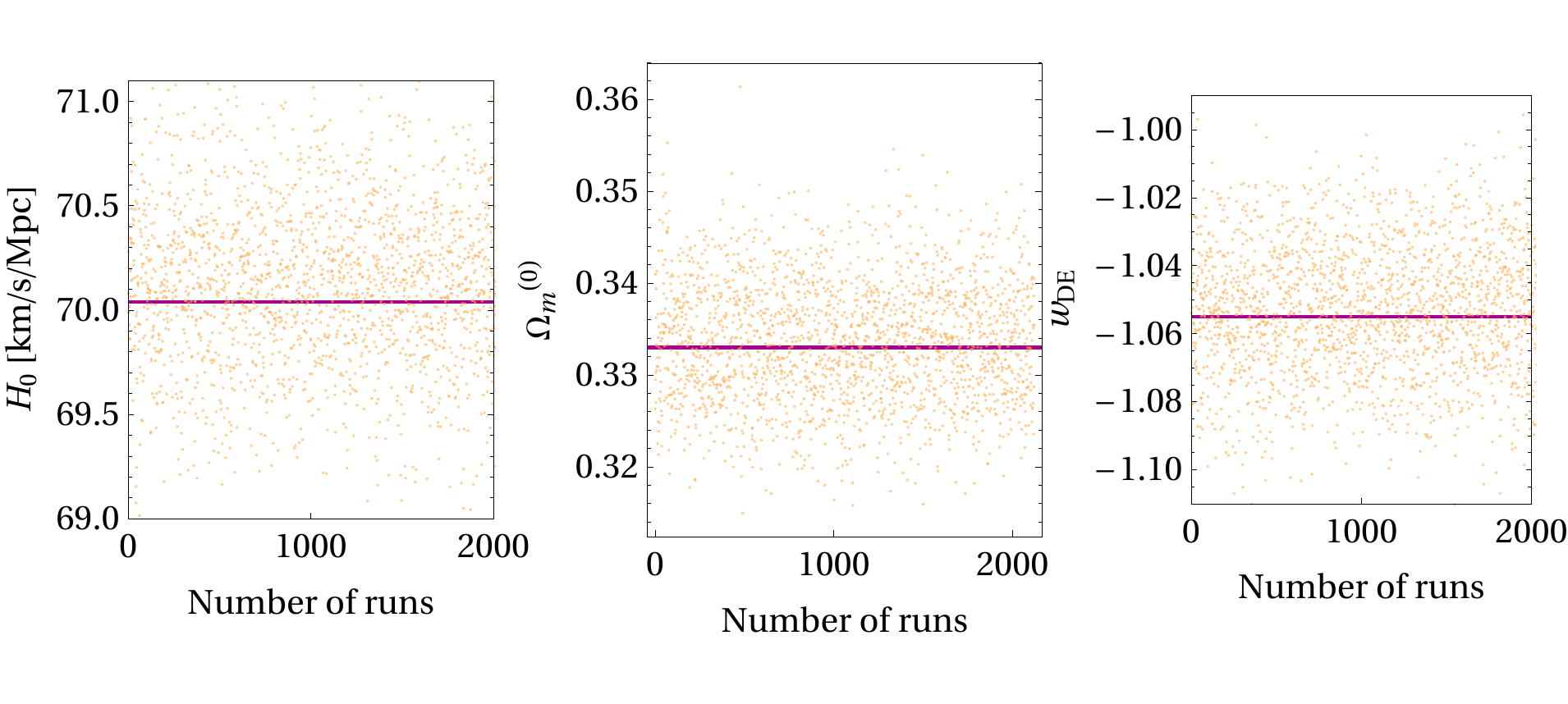}
    \caption{With $H(z)$+Pantheon+BAO.}
\end{subfigure}
    \caption{Figures (a), (b) and (c) illustrate the parameter values allowed by comparing  $H_{alg}(z)$ and $\mathcal{H}(z)$ using the Monte Carlo 
    method for OHD and its combination with SN1a dataset and Pantheon+BAO dataset. The solid line showcase their respective best-fit results. }
    \label{fig:back_params}
\end{figure}

To obtain the values of cosmological 
parameters that correspond to the best-fit form generated by 
the algorithm, denoted as $H_{alg}(z)$, we aim to minimize the 
sum of squared errors, defined as follows:
\begin{equation}
    \zeta: = \sum_i(H_{alg}(z_i) - \mathcal{H}(z_i))^2 
\end{equation}
where $z_i \in [0,1]$ and we divide the range of $z$ into bins of 
size $0.01$. Also, the prior distribution of the parameters are given as:
\begin{equation}
    H_0 \in \mathcal{U}[60,80] \,, \quad 
    \Omega_m^{(0)} \in \mathcal{U}[0.2,0.5] \,, \quad 
    w_{DE} \in \mathcal{U}[-1.5, -0.8] \,.
\end{equation}
By using the Markov Chain process as a sampling 
technique we obtain the distribution and best-fit of parameters: 
$\{H_0, \Omega_m^{(0)}, w_{DE}\}$. 
It is worth noting that our estimation of $H_0$ aligns 
with the findings of direct measurements from the distance-ladder technique, such as SH0ES ($H_0 = 73.04 \pm 1.04$ Km/s/Mpc) and other low-redshift observations such as the Megamaser Cosmology Project (MCP) 
\cite{mcp}, $H_0$ Lenses in COSMOGRAIL's Wellspring 
(H0LiCOW) \cite{holicow}. These observations measure $H_0 
= 73.9 \pm 3$ Km/s/Mpc, and $73.3^{+1.7}_{-1.8}$ Km/s/Mpc, respectively. 

Let us emphasize that the significant enhancement in 
the value of $H_0$ (as shown in fig.\,(\ref{fig:back_params})) and the 
effectively resolution of the Hubble 
tension can be attributed primarily to the phantom behavior of 
dark energy, rather than relying on the commonly speculated under-density 
of matter at low-$z$ \cite{low-matter}. This is due to the fact that in our 
estimations, the best-fit values of $\Omega_m^{(0)}$ are found to be around 
$0.35$ (see fig.\,(\ref{fig:back_params})). In fact, 
the enhancement in $H_0$ as well as $\Omega_m^{(0)}$ leads to an 
approximate $6\%$ increase in the total (local) matter density, 
given by $\Omega_m h^2$, compared to what is predicted from the 
Planck results. On the other hand, the large negative values of 
equation-of-state-parameter support the phantom-like nature of DE. 
In a nutshell, the significant 
level of discrepancy in the measurement of cosmological parameters 
between parametric and non-parametric methods at the background 
level is primarily attributed to the inherent biases present in 
models such as the $\Lambda$CDM. If these biases can be mitigated to 
some extent, the phantom phase, which may not necessarily be mild, 
aligns more favorably with the observed data. 

For the sake of verification, we have also checked the validity of our 
results, i.e. if the phantom-like behaviour is necessarily the reason 
for the enhancement of the Hubble constant value or is it just 
the artifact of the choosing $w$CDM template. In order verify this we
consider a more general as well as a theoretically 
motivated interacting DE-matter scenario which appears in a large 
class of modified gravity theories. In this scenario, the coupling 
takes the form of $Q\rho_m \dot{\phi}$ \cite{AT-book}, and which does not 
assume a constant equation-of-state-parameter for DE. 
In fact, in this case the DE equation of state depends on the coupling 
as well as on the matter density parameter and is evolving in nature. 
When compared with the dataset-2 obtained $H_{alg}(z)$ functional form 
(\ref{H_comb}), 
we have found that its corresponding best-fit of $w_{DE}$ is 
around $-1.087$, which is even slightly larger than what we have 
found earlier. This consistency in the results from two different 
cosmological frameworks with two different sets of background level data 
indicates that our results are not specific to a given framework. 
Moreover, our result also corroborate with ref.\,\cite{periv} 
where it was shown that in order to alleviate the tension the 
DE equation of state must reside in the deep phantom regime.

\subsection{Possible physical interpretations of the Hubble parameter form}

Let us now look for the conceptual implications of the class of theories 
to which the Hubble parameter expressions (\ref{H_alg}, \ref{H_alg_SN1a}, 
and \ref{H_comb}) may be more closely associated. While these derived 
expressions are entirely numerical in nature, we have also demonstrated 
their preference for the phantom-like characteristics of DE. This 
DE source can potentially originate from single(or multi-)field(s) cosmological 
scenarios, in various class of scalar-tensor equivalent 
modified gravity theories 
\cite{phantom, phantom2}. This characteristic can also appear in disformal coupling between baryonic 
and dark matter \cite{generic, generic2} which does not assume any 
extra degrees of freedom. However, identifying which scenario 
is more preferable to give rise to the algorithm assisted Hubble 
parameter form at the observational level poses a formidable challenge. 
Therefore, at this point we can only anticipate that the observed behavior 
of the Hubble parameter may emerge within some specific, well-defined 
cosmological scenarios. If it has to be stemmed out within the Einstein 
frame, there must be atleast two minimally-interacting scalar fields with 
the matter sector, whether they are in canonical or non-canonical form, 
such as one considers in the standard quintom scenarios 
\footnote{It is important to highlight that in \cite{sola-4, sola-5} it is shown that the phantom DE may emerge as an effective behavior originating from the quantum vacuum.}. On the other hand, the phantom nature can also manifest in non-minimally interacting scenarios (or within 
the Jordan frame), depending upon the chosen coupling(s) between the scalar field and gravity. Furthermore, in disformal coupling scenarios as discussed 
in \cite{generic}, it is possible to achieve phantom dark energy behavior 
when one of the fluids, like baryonic matter, adheres to the geodesic of the Jordan frame, while dark matter follows that of the Einstein frame, and 
through the disformal coupling it gives rise to phantom DE in the Jordan 
frame. The model-independent Hubble parameter form allows us to 
look for a more general cosmological scenario that can address 
current cosmological tensions. However, determining their consistency 
with the field equations relies entirely on the specific characteristics 
of DE. Given that both baryonic and cold dark matter evolve according to $(1+z)^3$, any segment of the observed $H_{alg}(z)$ that remains after 
subtracting this component can be attributed to the `effective' 
dark energy for the flat-universe i.e. 
\begin{equation}
    \rho_{_{DE}}(z) \equiv \frac{3 H_{alg}^2(z)}{8\pi G_{_N}} - \rho_m(z) \,, \quad \mbox{where} \quad 
    \rho_m(z) = \rho^{(0)}_m \, (1+z)^3 \,.
\end{equation}

It is also worth noting that the various 
``fitness levels'' or the $\chi^2$ involved in the final optimization 
process may correspond to specific cosmological scenarios, at least 
those closely approaching the optimal value. For instance, a particular 
fitness level might align with a specific cosmological model, like 
quintessence models. However, if it exceeds their fitness, it suggests 
the possibility of a better theoretical model that can more accurately 
fit the data.

\section{Linear growth rate of matter density perturbations}

Several recent low-$z$ observations of the large-scale structure 
allow us to figure out the extent of matter density clustering 
in the universe. In order to analyze it, we utilize the same 
optimization algorithm to analyze data pertaining to matter 
perturbation, specifically focusing on redshift-space distortions (RSD).
For this dataset, we have used the compilation of 
$f \sigma_8(z)$ observations, where $f$ is the growth factor of matter perturbations, and $\sigma_8$ is the amplitude of power spectrum in $8h^{-1}$Mpc ] \cite{KPS-brs}, and is related to the Power spectrum $P(k)$ via \cite{AT-book}
\begin{equation}
    \sigma_8^2 = \frac{1}{2\pi} \int W_s^2 k^2 P(k) dk
\end{equation}
where $k$ is the comoving wavenumber, and $W_s$ is the window function. 
We consider the Growth-Gold compilation of $f\sigma_8$ measurements 
obtained from various galaxy surveys within the redshift interval of 
$z \in [0.02, 1.94]$ \cite{growthdata}. 
The main reason for opting for this specific subset of data is due 
to its uncontaminated nature, lack of anomalies and 
widely usage (see refs. \cite{sunny,perivola,nunes}).  
The $\chi^2$ for the same is defined as: 
\begin{equation}
    \chi^2 := V^i \, C_{ij}^{\, -1} \, V^j \,, \quad \mbox{where} \quad 
    V \equiv f \sigma_8(z)- [f \sigma_8(z)]_{\text{alg}} \,,
\end{equation}
where $[f \sigma_8(z)]_{\text{alg}}$ represents the best-fit of the 
algorithm, and $C_{ij}$ is the covariance matrix between different data points. In line with the background analysis, we have carried out multiple 
simulations using the identical procedure applied to the cosmological background level. Furthermore, we have also examined various 
initial values to determine whether the resulting fit exhibit any 
differences with each other. The best-fit function and it corresponding 
minimized $\chi^2$ is given as
\begin{equation} \label{fs8-fit}
    [f\sigma_8]_{alg}(z) = 0.537 e^{0.159 z} -10^{-5} z^4+0.098z^3-0.359 z^2+0.216 z -0.163 \,, \quad \mbox{with} \quad \chi^2 = 11.91 \,,
\end{equation}
\begin{figure}
    \centering
    \includegraphics[height=2.7in, width=4.2in]{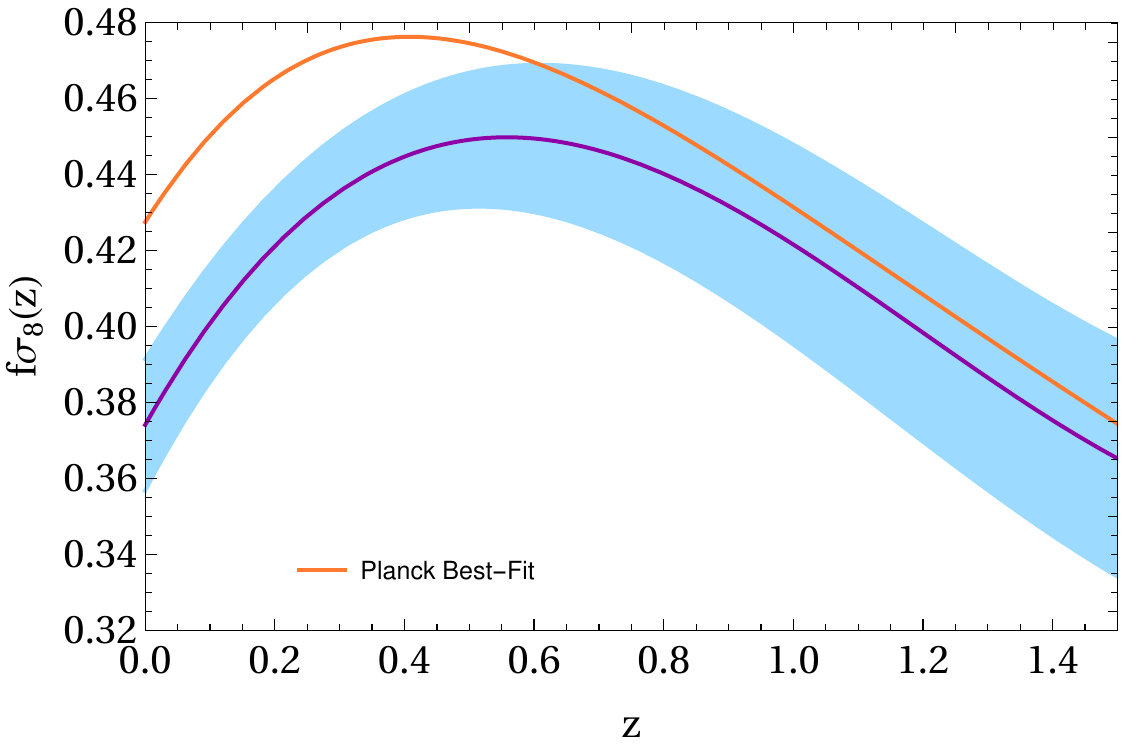}
    \caption{\small Optimized growth of matter density clustering with redshift $z$ within the $1\sigma$ limit, and Planck's proposed trajectory. Both trajectories merge together at the high-redshifts. 
    }
    \label{fig:fs8}
\end{figure}
For the obtained $[f\sigma_8]_{alg}(z)$ fit, we depict its evolutionary 
profile up to $1 \sigma$ level in fig.\,(\ref{fig:fs8}). Let us note that 
at the present epoch, Eq.\,(\ref{fs8-fit}) gives
\begin{equation} \label{fs80}
[f\sigma_8]_{alg}(0)  =  0.374 \pm 0.017 \,,
\end{equation}
which is significantly lower (a level $>2 \sigma$) than the Planck result of 
$0.474 \pm 0.015$. Since in obtaining the result (\ref{fs80}) no parametric 
or functional form was assumed, and it still shows a significant level 
of tension with the Planck's result, it certainly lead to the conclusion 
that the discrepancy exists at the level of the observations. As we have 
observed that the discrepancy in $H_0$ measurements at the background 
level is associated with the predominantly phantom-like nature of DE, 
one may ask: \textit{whether the same parameter 
reflects a similar discrepancy at the perturbative level?} In order 
to verify this we will proceed with the same procedure of parameter 
estimations.

\subsection{Growth rate parameter estimations}

In order to comprehend the implications of fig.\,(\ref{fig:fs8}) in 
terms of cosmological parameters associated to the growth of matter perturbations, such as $\Omega_m^{(0)}$, $\sigma_8^{(0)}$ and $w_{_{DE}}$, 
we re-consider the flat-$w$CDM model. For the latter, 
the equation of motion of matter density contrast 
is given by \cite{bias}
\begin{equation} \label{delta-m}
    \delta_m''+\frac{1}{2}\left[1- 3 (1-\Omega_m(a))w_{_{DE}} \right] \delta_m' = \frac{3}{2} \Omega_m \, \delta_m \,,
\end{equation}
where $'$ denotes the derivative with respect to $\log(a)$ and $a$ is the scale factor. In general form, the analytical solution 
of the matter density contrast $\delta_m$ can be found as
\begin{equation} \label{delta}
\frac{\delta_m}a = \,
 _2 F_1\left(\frac{w_{_{DE}} - 1}{2w_{_{DE}}},\frac{-1}{3 w_{_{DE}}};1-\frac{5}{6 w_{_{DE}}};
 a^{-3 w_{_{DE}}} \left(1-\frac{1}{\Omega_m^{(0)} }\right)\right)  \,,
 \end{equation}
where $_2\,F_1$ is the Hypergeometric function.
Using this one can calculate the theoretical growth rate as
\begin{equation} \label{fs8}
    f \sigma_8(z) = f(z)\, \sigma_8^{(0)} \, 
    \frac{\delta^{(m)}(z)}{\delta^{(m)}(0)} \,.
\end{equation}
Here again, we adopt the same approach to statistically compare $f 
\sigma_8(z)$ with Eq.\,(\ref{fs8}). In particular we try to minimize the
squared-difference between the $[f \sigma_8]_{alg}(z)$ and 
$f \sigma_8(z)$. The estimated values are given as follows \footnote{The 
obtained minimized $\chi^2$ value is better than 
that of the corresponding $\Lambda$CDM model and the $w$CDM model. The primary objective of the estimations is to illustrate the potential range of values achievable for the fitting (\ref{fs8-fit}).}:
\begin{eqnarray} \label{estimates}
     w_{_{DE}} = -1.596 \pm 0.099 \,, \quad \Omega_m^{(0)} = 0.338 \pm 0.083  \,, \quad 
    \sigma_8^{(0)} = 0.795 \pm 0.072 \,.
\end{eqnarray}
\begin{figure}
    \centering
    \includegraphics[height=6.5cm,width=10cm]{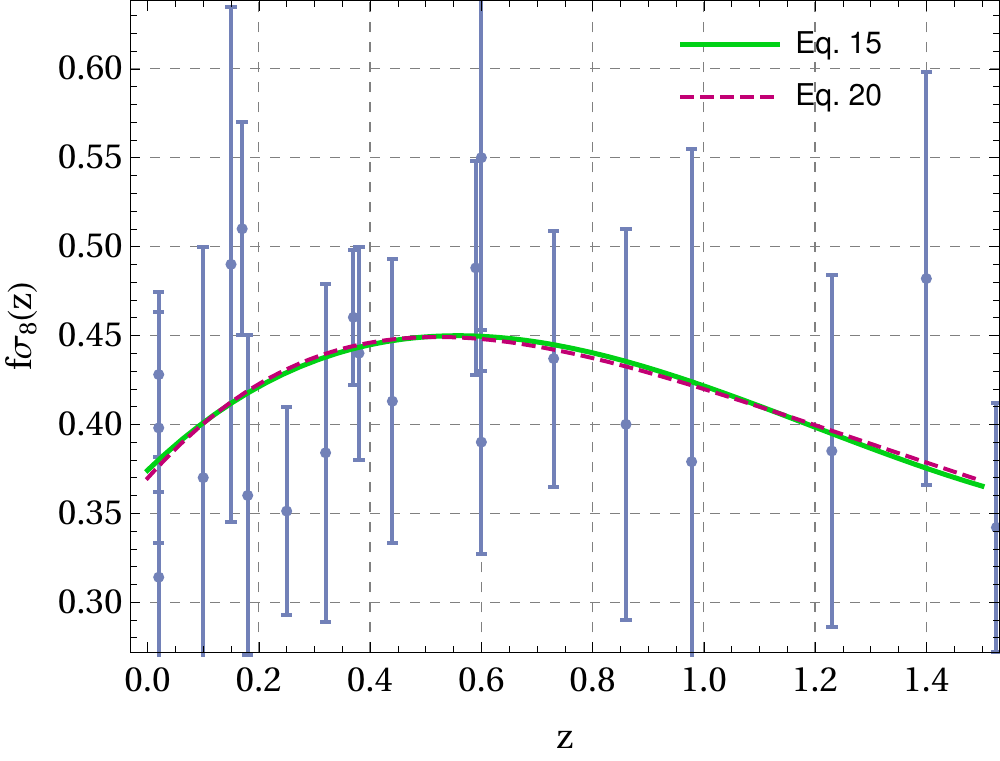}
    \caption{The figure shows the degree of accuracy in fitting the curve of the $w$CDM model, using the estimated parameter values (\ref{estimates}), with the algorithmically predicted profile of $f\sigma_8(z)$ (\ref{fs8-fit}). The solid line represents the predicted profile of $f\sigma_8(z)$ generated by the algorithm, while the dashed line represents the fitting of the $w$CDM model to the aforementioned prediction. The nearly identical evolutionary patterns of these two curves suggest that our estimates closely match the form generated by the algorithm (\ref{fs8-fit}).} The corresponding Planck's 
    $\Lambda$CDM curve is shown in fig.\,(\ref{fig:fs8}).
    \label{fig:fs8_wcdm-alg}
\end{figure}

Here also, we see that the equation of state for DE favours its phantom 
nature by leaning slightly towards lower values ($<-1$). On the other 
hand, the value of $\sigma_8^{(0)}$ is significantly higher than what was 
predicted by low-redshift observations such as KiDS-450 \cite{kids450} and KiDS-1000 \cite{kids1000}. As 
already mentioned that the KiDS-450 estimate of $\sigma_8^{(0)} = 0.745 
\pm 0.039$ exhibits a tension of more than
$2 \sigma$ with Planck TT,TE,EE+lowE+lensing estimate 
$\sigma_8^{(0)} = 0.811 \pm 0.006$ \cite{planck18}. Notably, 
our estimate on $\sigma_8^{(0)}$ does not show any 
tension with Planck's $\sigma_8^{(0)}$ result and is in 
agreement with the latter. In fig.\,(\ref{fig:fs8_wcdm-alg}) 
we depict the accuracy of the best-fit values 
obtained from the parametric estimations 
(\ref{estimates}) in relation to the algorithm-predicted 
$f\sigma_8(z)$ profile. The figure demonstrates that the 
profiles of both the estimations and the algorithm 
prediction are in alignment, indicating that the 
estimations (\ref{estimates}) are reasonably accurate and 
exhibit a significant level of goodness of fit. 

\subsection{$S_8$ Constraints}
\begin{figure}
    \centering
    \includegraphics[height=2.9in, width=5.2in]{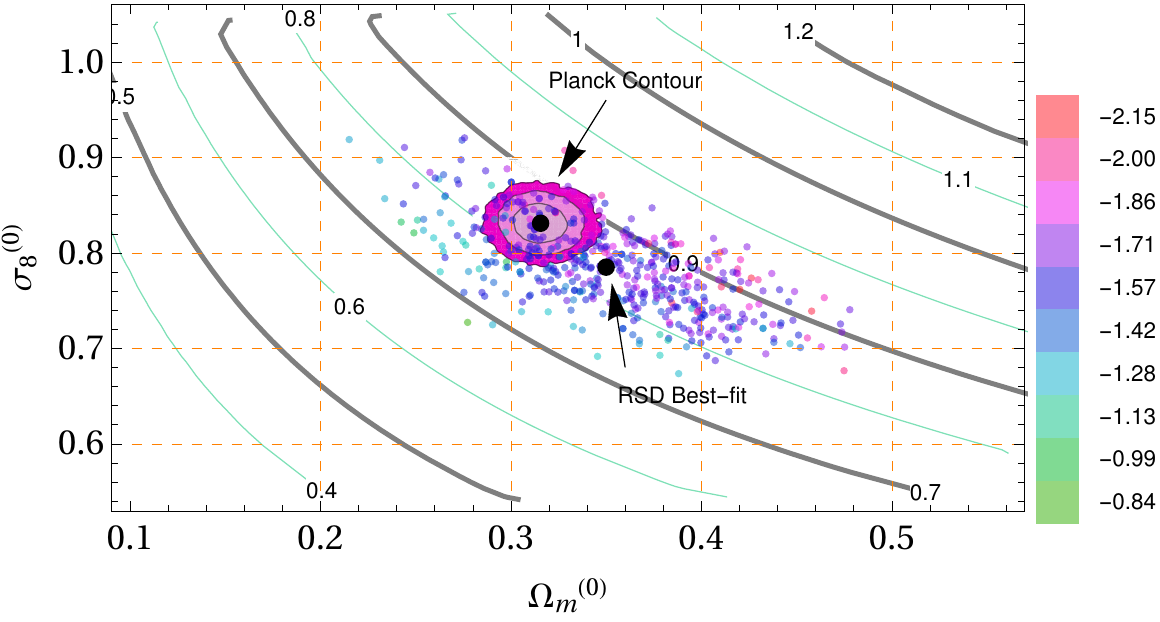}
    \caption{The figure illustrates the parametric region between $\Omega_m^{(0)}$ and $\sigma_8^{(0)}$, which is allowed based on the Monte Carlo method. The Monte Carlo method generates values distributed normally around the best-fit values (\ref{estimates}). The solid region represents the Planck allowed region within a $3\sigma$ range. The curved lines in the plot correspond to different values of $S_8^{(0)}$, and the values are indicated on each curve. The color bar on the right side of the figure provides the corresponding values of the equation of state for dark energy for each colored point shown in the plot.}
    \label{fig:enter-label}
\end{figure}
The weighted magnitude of matter density perturbations ($S_8$) 
captures the degeneracy between $\Omega_m^{(0)}$ and $\sigma_8^{(0)}$ and 
is formulated as
\cite{sunny}:
\begin{equation} \label{S8}
S_8^{\,(0)}  = \sigma_8^{\,(0)} \, \sqrt{\frac{\Omega_m^{(0)}}{0.3}} \,.
\end{equation}
In the framework of the $\Lambda$CDM model, measurements of CMB anisotropy by the Planck 2018 have yielded $S_8^{\,(0)} = 0.834 \pm 0.016$ 
\cite{planck18}. In contrast, a number of surveys of RSD consistently suggests $S_8^{\,(0)}$ values that tend to be 
lower than those inferred from CMB measurements, falling within the range of $[0.703,0.782]$.
However, when using the estimates (\ref{estimates}) in Eq.\,(\ref{S8}), 
and using the error-propagation technique, we find
\begin{equation} \label{S8-est}
    S_8^{\,(0)} = 0.833 \pm 0.188 \,,
\end{equation}
for the RSD dataset. This indicates a notable increase in the value of $S_8^{(0)}$ (although with a considerable level of uncertainty) and approaches the estimation by Planck TT,TE,EE+lowE of $S_8^{\,(0)} = 0.834 \pm 0.016$ \cite{planck18}. 
This suggests that if not be due to systematics, any disagreement or 
tension between the high and low redshift data might be due to the choice of the cosmological model which is used to describe the universe at 
late times.

\section{Joint background and perturbative level evolution}

As we have earlier shown that in order to address the 
tension at both the background and linear perturbative 
levels, it is important to deviate from the standard 
$\Lambda$CDM model towards a more phantom-like behavior. 
It is also important to note that both the estimates 
obtained from background and perturbative level data for 
parameters such as $\Omega_m^{(0)}$ and $w_{DE}$ are 
consistent with each other. This indicates that both sets 
of data align with each other and allow us to find a 
unified trajectory for the evolutionary history of the 
universe, accounting for both the growth of large-scale structure and the rate of expansion. Therefore, by using the Eqs.\,(\ref{H_alg}) and (\ref{fs8-fit}), we can analyze how the 
quantities $[f \sigma_8]_{alg}(z)$ and $H_{alg}(z)$ change together over a range of redshifts. This will 
give us the hint for the possible evolutionary profile of the universe which is required for reducing or rather resolving the existing tensions between the measurements. 
It will also allow us to examine the joint 
background and perturbative evolution of the universe without being limited by any constraints on the parameter. 

\begin{figure}[t]
    \centering
    \includegraphics{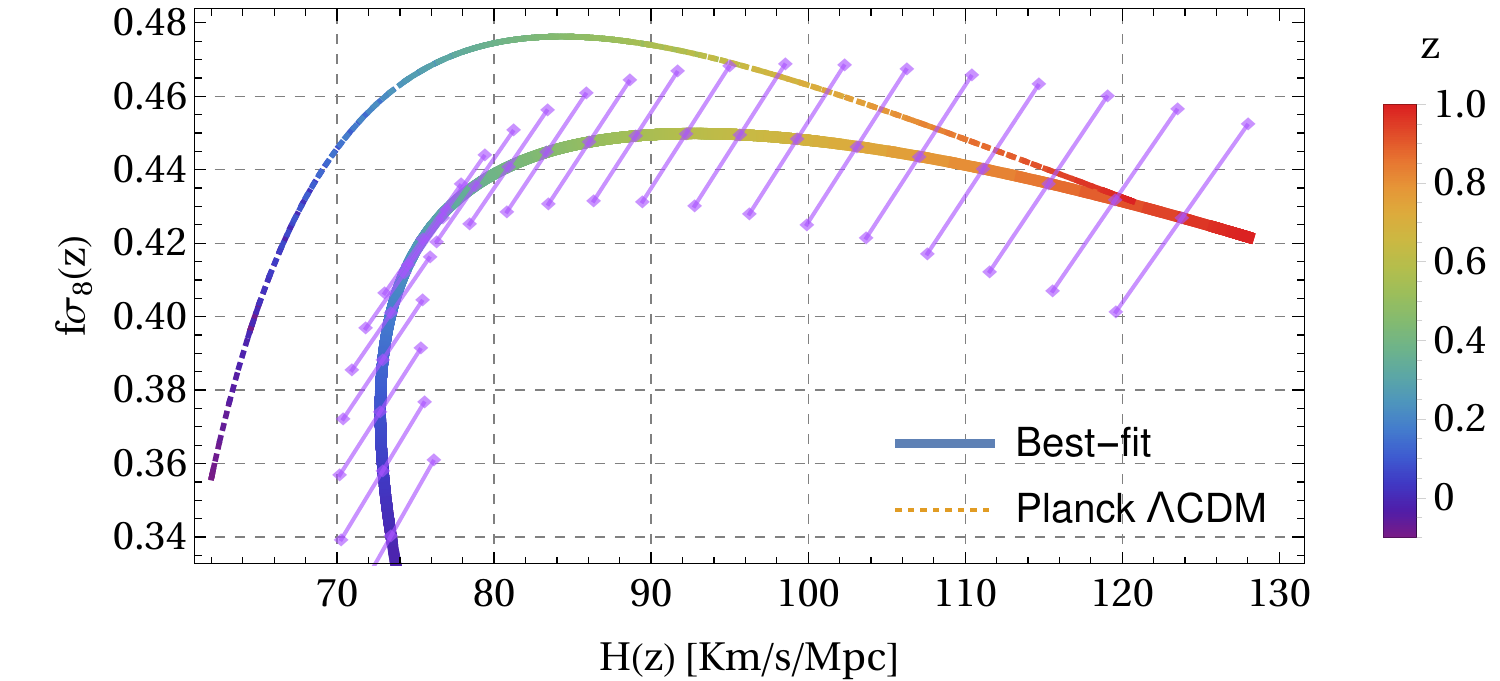}
    \caption{\small Evolution of $f\sigma_8(z)$ is shown versus $H(z)$ for $z \in[0,1.]$ for Planck and for our case.}
    \label{hubble-fs8}
\end{figure}

The obtained profile is shown in fig.\,(\ref{hubble-fs8}) in which we have 
also depicted the evolution of linear perturbation with the background 
expansion by utilizing the Planck TT,TE,EE+lowE+lensing best-fits 
(dotted curve) with $\Omega^{(0)}_m=0.315$ and $\sigma_8^{(0)} = 0.8111$ 
\cite{planck18}. 
The figure illustrates that both trajectories (fitted one and that correspond 
to the Planck's best-fit) have followed similar behaviour in the past. 
However, a noticeable deviation from each other is observed at $z \leq 0.4$. 
Also, the growth rate of matter perturbations for the Planck tends to be less suppressed than our case, and therefore reaches a peak value that is higher than what is predicted by our analysis almost at the same redshift value. 

As we have seen that the joint evolution of $f\sigma_8(z)$ and $H(z)$ is 
unique in the sense that it corresponds to those cosmological parametric 
values at the current epoch which relieves the tensions. This is in 
contrast to most scenarios where the particular correlation between 
the $H_0$ and $\sigma_8^{(0)}$ estimates for a given model tends to worsen the 
one while solving the other.

To assess the compatibility between the two observables, 
namely $H(z)$ and $f \sigma_8(z)$, and to determine the profile suggested 
by the background data for $f \sigma_8(z)$, one needs to know the background 
parameters, such as $\Omega^{(0)}_m$ and $w_{_{DE}}$, to use in 
Eq.,(\ref{delta-m}). 
By utilizing the best-fitting values of the background parameters obtained 
in section,(\ref{background}) in Eq.,(\ref{fs8}), and comparing it with the 
$f \sigma_8(z)$ profile derived from the RSD data estimations (\ref{estimates}), 
we find $\sigma_8^{(0)}\simeq 0.77$. This value is in agreement with the 
estimated value given in Eq.,(\ref{estimates}). Consequently, our background 
and perturbative-level analyses align with each other, indicating 
consistency in our results. 

In order to assess whether the $f \sigma_8(z)$ profiles 
obtained from the background data estimations are compatible with the one 
that is obtained using the RSD dataset, we use a simple technique that 
calculates the area between any two functional profiles. The area which 
denotes the divergence between two profiles is given by 
$A = \int_z |f(z)- g(z)|dz$. Hence, larger the area the less compatibility 
between two profiles, or vice-versa. After applying this technique, we 
have found that the value of $A$ for the estimations (\ref{estimates}) and 
the background estimations (\ref{background}) is approximately $\mathcal{O}
(10^{-3})$, whereas when compared with the $\Lambda$CDM estimations, $A$ 
is of the order of $\mathcal{O}(10^{-2})$. The comparatively large 
compatibility between our background and linear perturbative-level estimations 
agrees to the fact that there is a significant level of deviation from the $\Lambda$CDM model towards the phantom.
 
\section{Conclusion}

We have conducted two comprehensive independent analyses to identify the 
reasons behind two tensions related to parameters $H_0$ and $\sigma_8^{(0)}$ 
using a metaheuristic optimization technique. 
To determine the necessary 
requirement(s) that align(s) most favorably with the optimized form of the 
observables obtained by our algorithm, we have chosen the $w$CDM 
model for simplicity. However, we have also demonstrated that the 
outcomes are also consistent with interacting dark energy scenarios.
Notable, our cosmological model-agnostic 
findings have demonstrated that the phantom nature of a dynamical DE is 
required to relax both of the tensions. We have also shown that in order 
to tackle both the tensions together a specific profile of a trajectory 
between $H(z)-f\sigma_8$ is required.

Regarding the background evolution, we have obtained the fitting
using the metaheuristic optimization algorithm for the
$H(z)$ for two separate cases to figure out if the results are indeed 
pointing towards the same physics or not. Hence, in 
first analysis we take distance-ladder measurements and its combination 
with SN1a dataset, and in the second case we take the combination of 
the former with full Pantheon+ and BAO dataset. After obtaining the 
optimized functional form of $H_{alg}(z)$ from the simulation, 
we have then searched for the corresponding cosmological scenario, 
which can or at least try to resemble it. For the purpose of 
estimation, we choose $w$CDM model, and with multiple simulations, 
we have obtained an average values for the same, which correspond to the optimized fitting
profiles of $H(z)$. We have explicitly shown that because of the
natural emergence of the phantom nature of DE the $H_0$ tension is
relaxed/reduced in CCH and its combination with SN1a dataset and 
BAO dataset. Here we also want to 
mention that for the CC+SN1a+BAO dataset the 
observed mild phantom behavior indicated by the dark energy equation of state, denoted as $w_{DE} = -1.087$, is primarily a result of fixing the sound horizon to the best-fit value of the $\Lambda$CDM model for the BAO dataset. As a result, this significant biasing effect effectively limits the degree to which the CC+SN1a dataset can penetrate into the phantom regime. The extent to which phantom 
nature can exhibit without the BAO dataset has been explicitly demonstrated by our use of the CC+SN1a dataset in the case of dataset-1. Moreover, 
our overall conclusion is in contrast to the point of view 
that the tension might be due to the low-matter density in the 
universe. This is due to the fact that in all different set of 
combinations of dataset, we have found 
$\Omega_m^{(0)} \in [0.33,0.35]$. Since the corresponding results 
agree with both sets of data thereby establishing their reliability, 
while having the potential to alleviate/reduce the $H_0$ tension. 
Here it is also important to mention that late-time 
modifications or considering the alternatives of $\Lambda$CDM are necessary 
due to the fact that the early-time possible resolutions for tackling 
the Hubble tension suffers with various issues and does not fully 
resolve the tension \cite{sunny1}. 

As to the linear growth of matter density perturbations, we have carried out
a similar procedure for finding out the cosmological parameters using
RSD dataset. Here also, we have shown that the corresponding optimized
fitting surpasses the fitting of the $\Lambda$CDM by a significant margin.
We have then obtained the corresponding parameters using the latter and
it strongly supports the phantom nature of DE. It is also in tune with the
large $\sigma_8^{(0)}$ value, which eventually reduces the tension.
Furthermore, by using the obtained constraints, we have constrained
the $S_8^{(0)}$ parameter and showed that its best-fit also lies 
towards the Planck's estimate. We have also shown that our results 
obtained using RSD dataset are compatible with the background ones. 
In summary, we have shown that both at the background and linear 
perturbative levels, the tensions can be alleviated if one chooses a suitable candidate for DE which exhibits a phantom nature at late times.

Let us emphasise that our analysis, which involves multiple data sets and their combinations, consistently demonstrates a better fit when compared to the $\Lambda$CDM model. This suggests the potential necessity for DE to exhibit phantom-like behaviour. Furthermore, aligning the model-independent observables with two distinct cosmological templates strengthens our argument. This approach is better in the sense that it allows us to obtain even small features of the observables in the data without relying on initial model-dependent assumptions.

There are still some unanswered questions that remain: (i) What insights can the optimised fitting provide regarding the interaction between dark energy (DE) and matter? (ii) If a model is capable of reproducing the corresponding results, will it exhibit stability? (iii) What will 
happen if the sound horizon is not fixed 
in prior? We are looking forward to answering these questions and trying to report on them in the near future.

\section*{Acknowledgement}

We thank Savvas Nesseris, Eoin O. Colgain, and Maurice VanPutten for their fruitful discussions on the draft. 
We also thank the anonymous reviewer for careful reading of our manuscript and giving many insightful comments and suggestions. The work of MRG is supported by DST,
Government of India under the Grant Agreement number
IF18-PH-228 (Inspire Faculty Award). The work of MRG and MS is supported by Science and Engineering Research Board (SERB), DST, Government of India under the Grant Agreement number CRG/2022/004120 (Core Research Grant). MS is also partially supported by the Ministry of Education and Science of the Republic of Kazakhstan, Grant
No. AP14870191 and CAS President's International Fellowship Initiative(PIFI).


\begin{thebibliography}{99}
%
\bibitem{AT-book}
L. Amendola and S. Tsujikawa, {\em Dark Energy: Theory and Observations}, 
Cambridge University Press, United Kingdom (2010).
%
\bibitem{planck18}
N. Aghanim et. al., {\em Planck 2018 results. VI. Cosmological parameters}, Astron. Astrophys. {\bf 641} (2020) A6, arXiv:\,1807.06209 [astro-ph.CO].
%
\bibitem{riess}
A. G. Riess et. al., {\em A 2.4\% Determination of the Local Value of the Hubble Constant}, Astrophys. J., {\bf 826}(1) (2016) 56, arXiv:\,1604.01424 [astro-ph.CO]. 
%
\bibitem{holicow}
K. C. Wong et. al., {\em H0LiCOW \textendash{} XIII. A 2.4 per cent measurement of H0 from lensed quasars: 5.3\ensuremath{\sigma} tension between early- and late-Universe probes}, Mon. Not. Roy. Astron. Soc., {\bf 498}(1) (2020) 1420-1439, arXiv:\,1907.04869 [astro-ph.CO].
%
\bibitem{kids450}
H. Hildebrandt et. al., {\em KiDS-450: Cosmological parameter constraints from tomographic weak gravitational lensing}, Mon. Not. Roy. Astron. Soc., {\bf 465} (2017) 1454, arXiv:\,1606.05338 [astro-ph.CO].
%
\bibitem{kids1000}
C. Heymans et. al., {\em KiDS-1000 Cosmology: Multi-probe weak gravitational lensing and spectroscopic galaxy clustering constraints}, Astron. Astrophys. {\bf 646} 
(2021) A140, arXiv:\,2007.15632 [astro-ph.CO].
%
\bibitem{elen}
E. Di Valentino, et. al., {\em Cosmology Intertwined III: $f \sigma_8$ and $S_8$}, Astropart. Phys. {\bf 131} (2021) 102604, arXiv:\,2008.11285 [astro-ph.CO].
%
\bibitem{val-rev}
E. Di Valentino et al., {\em In the realm of the Hubble tension—a review of solutions}, Class. Quant. Grav. 
{\bf 38} (2021) 15, 153001, 
arXiv:\,2103.01183 [astro-ph.CO].
%
\bibitem{running-H0sig8}
J. Sol\`a Peracaula, A. Gomez-Valent, and J. de C. Perez, C. M.-Pulido, {\em Running vacuum in the Universe: phenomenological status in light of the latest observations, and its impact on the $\sigma_8$ and 
$H_0$ tensions}, Universe {\bf 9} (2023) 6, 262, arXiv:\,2304.11157 [astro-ph.CO].
%
\bibitem{dain-hTension}
M. G. Dainotti, G. Bargiacchi, M. Bogdan, 
S. Capozziello, S. Nagataki, {\em Reduced uncertainties up to $43\%$ on the Hubble constant and the matter density with the SNe Ia with a new statistical analysis}, arXiv:\,2303.06974 [astro-ph.CO].
%
\bibitem{dain-hTension2}
G. Bargiacchi, M.G. Dainotti, S. Capozziello, 
{\em Tensions with the flat $\boldsymbol{\Lambda}$CDM model from high-redshift cosmography}, arXiv:\,2307.15359 [astro-ph.CO].
%
\bibitem{nunes}
R. D'Agostino, R. C. Nunes, {\em Cosmographic view on the H0 and \ensuremath{\sigma}8 tensions}, 
Phys. Rev. {\bf D} {\bf 108(2)} (2023) 023523, 
arXiv:\,2307.13464[astro-ph.CO].
%
\bibitem{cpsingh}
J. de C. Perez, J. Sol\`a Peracaula, C.P. Singh, 
{\em Running vacuum in Brans-Dicke theory: a possible cure for the $\sigma_8$ and $H_0$ tensions}, arXiv:\,2302.04807[astro-ph.CO].
%
\bibitem{huterer}
N. Nguyen, D. Huterer, and Y. Wen, {\em 
Evidence for suppression of structure growth in the concordance cosmological model}, arXiv:\,2302.01331[astro-ph.CO].
%
\bibitem{amendola}
B. J. Barros, and L. Amendola, and T. Barreiro, and N. J. Nunes, {\em Coupled quintessence with a $\Lambda$CDM background: removing the $\sigma_8$ tension}, JCAP {\bf 01} (2019) 007, arXiv:\,1802.09216 [astro-ph.CO].
%
\bibitem{sujikawa}
J. B. Jiménez, D. Bettoni, D. Figueruelo, F. A. T. Pannia, and S. Tsujikawa, {\em Probing elastic interactions in the dark sector and the role of S8}, Phys. Rev. {\bf D} {\bf 104}(10) (2021) 103503, arXiv:\,2106.11222 [astro-ph.CO].
%
\bibitem{pandey}
K. L. Pandey, T. Karwal, and S. Das, {\em Alleviating the $H_0$ and $\sigma_8$ anomalies with a decaying dark matter model}, JCAP {\bf 07} (2020) 026, 
arXiv:\,1902.10636 [astro-ph.CO].
%
\bibitem{trodden}
M. C. Carrillo, Mariana, Q. Liang, and J. Sakstein, and M. Trodden, {\em Neutrino-Assisted Early Dark Energy is a Natural Resolution of the Hubble Tension}, arXiv:\,2302.09091 [astro-ph.CO].
%
\bibitem{trodden2}
J. Sakstein, and M. Trodden, {\em Early Dark Energy from Massive Neutrinos as a Natural Resolution of the Hubble Tension}, Phys. Rev. Lett. {\bf 124}(16) (2020) 161301, arXiv:\,1911.11760 [astro-ph.CO].
%
\bibitem{sunny1}
S. Vagnozzi, {\em Seven hints that early-time new physics alone is 
not sufficient to solve the Hubble tension}, Universe {\bf 9} (2023) 393, 
arXiv:\,2308.16628 [astro-ph.CO].
%
\bibitem{periv}
G. Alestas, L. Kazantzidis, and L. Perivolaropoulos, 
{\em $H_0$ tension, phantom dark energy, and cosmological parameter degeneracies}, Phys. Rev. {\bf D} {\bf 101}(12) (2020) 123516, 
arXiv:\,2004.08363 [astro-ph.CO].
%
\bibitem{generic}
S. A. Adil, M. R. Gangopadhyay, M. Sami, and M. K. Sharma, {\em Late-time acceleration due to a generic modification of gravity and the Hubble tension}, Phys. Rev. {\bf D} {\bf 104}(10) (2021) 103534, arXiv:\,2106.03093 [astro-ph.CO].
%
\bibitem{generic2}
M. R. Gangopadhyay, S. K. J. Pacif, M. Sami, and M. K. Sharma, {\em Generic Modification of Gravity, Late Time Acceleration and Hubble Tension}, Universe {\bf 9}(2) (2023) 83, arXiv:\,2211.12041 [gr-qc].
%
\bibitem{mont-htension}
G. Montani, M. De Angelis, F. Bombacigno, N. Carlevaro, {\em Metric $f(R)$ gravity with dynamical dark energy as a paradigm for the Hubble Tension}, 
arXiv:\,2306.11101 [gr-qc]
%
\bibitem{sola-1}
J. Sol\`a Peracaula, A. G\'omez-Valent, J. de C. Perez, 
C. Moreno-Pulido, {\em Running vacuum against the $H_0$ 
and $\sigma_8$ tensions}, EPL {\bf 134(1)} (2021) 19001, 
arXiv:\,2102.12758 [astro-ph.CO].
%
\bibitem{sola-2}
J. Sol\`a Peracaula, A. G\'omez-Valent, J. de C. Perez, 
C. Moreno-Pulido, {\em Running Vacuum in the Universe: Phenomenological Status in Light of the Latest Observations, and Its Impact on the \ensuremath{\sigma}$_{8}$ and H$_{0}$ Tensions}, 
Universe {\bf 9(6)} (2023) 262, 
arXiv:\,2304.11157 [astro-ph.CO].
%
\bibitem{sola-3} 
C. Moreno-Pulido, and J. Sol\`a Peracaula, {\em Renormalizing the vacuum energy in cosmological spacetime: implications for the cosmological constant problem}, Eur. Phys. J. {\bf C} {\bf 82(6)} (2022) 551, 
arXiv:\,2201.05827 [gr-qc].
%
\bibitem{sola-4}
C. Moreno-Pulido, and J. Sol\`a Peracaula, 
{\em Equation of state of the running vacuum}, 
Eur. Phys. J. {\bf C} {\bf 82(12)} (2022) 137, 
arXiv:\,2207.07111 [gr-qc].
%
\bibitem{sola-5}
C. Moreno-Pulido, J. Sol\`a Peracaula and S. Cheraghchi, 
{\em Running vacuum in QFT in FLRW spacetime: the dynamics of $\rho _{\textrm{vac}}(H)$ from the quantized matter fields}, Eur. Phys. J. {\bf C} {\bf 83(7)} (2023) 
637, arXiv:\,2301.05205 [gr-qc].
%
\bibitem{silk}
E. D. Valentino, A. Melchiorri, and J. Silk, {\em Planck evidence for a closed Universe and a possible crisis for cosmology}, Nature Astron. {\bf 4}(2) (2019) 196-203, arXiv:\,1911.02087 [astro-ph.CO].
%
%
\bibitem{abdalla}
E. Abdalla, {\em Cosmology intertwined: A review of the particle physics, astrophysics, and cosmology associated with the cosmological tensions and anomalies}, JHEAp 
{\bf 34} (2022) 49-211, arXiv:\,2203.06142 [astro-ph.CO]
%
\bibitem{elenora}
E, D. Valentino et. al., {\em Cosmology Intertwined III: $f \sigma_8$ and $S_8$}, Astropart. Phys. {\bf 131} (2021) 102604, arXiv:\,2008.11285 [astro-ph.CO]. 
%
%
\bibitem{sigma8drag}
V. Poulin, J. L. Bernal, E. Kovetz, and M. Kamionkowski, {\em The Sigma-8 Tension is a Drag}, arXiv:\,2209.06217 [astro-ph.CO].
%
\bibitem{pogosian}
K. Jedamzik, and L. Pogosian, {\em Relieving the Hubble tension with primordial magnetic fields}, Phys. Rev. Lett. {\bf 125}(18) (2020) 181302, arXiv:\,2004.09487 [astro-ph.CO].
%
\bibitem{perivola}
G. Alestas, and L. Perivolaropoulos, {\em Late-time approaches to the Hubble tension deforming H(z), worsen the growth tension}, Mon. Not. Roy. Astron. Soc. {\bf 504}(3) (2021) 3956-3962, arXiv:\,2103.04045 [astro-ph.CO].
%
\bibitem{supratik}
R. Shah, A. Bhaumik, P. Mukherjee, S. Pal, {\em A thorough investigation of the prospects of eLISA in addressing the Hubble tension: Fisher Forecast, MCMC and Machine Learning}, arXiv:\,2301.12708 [astro-ph.CO].
%
\bibitem{neserris2}
R. Arjona, and S. Nesseris, {\em What can Machine Learning tell us about the background expansion of the Universe?}, Phys. Rev. {\bf D} {\bf 101}(12) 
(2020) 123525, arXiv:\,1910.01529 [astro-ph.CO].
%
\bibitem{neserris3}
R. Arjona, and S. Nesseris, {\em Hints of dark energy anisotropic stress using Machine Learning}, JCAP {\bf 11} (2020) 042, arXiv:\,2001.11420 [astro-ph.CO].
%
\bibitem{colgain}
E. O. Colgáin, M. H.P.M. van Putten, H. Yavartanoo, 
{\em de Sitter Swampland, $H_0$ tension \textbackslash{}\& observation}, Phys. Lett. {\bf B} {\bf 793} (2019) 126--129, arXiv:\,1807.07451 [astro-ph.CO].
%
\bibitem{GA1}
S. Mirjalili, and S. Mirjalili,  {\em Genetic algorithm. Evolutionary Algorithms and Neural Networks: Theory and Applications}, 43-55, (2019).
%
\bibitem{GA2}
J. H. Holland, {\em Genetic algorithms} Scientific american, {\bf 267(1)} (1992) 66-73.
%
\bibitem{GA3}
O. Kramer, and O. Kramer {\em Genetic algorithms} 
Springer International Publishing, 2017.
%
\bibitem{GA4}
J. Liesenborgs, S. De Rijcke, H. Dejonghe, 
{\em A genetic algorithm for the non-parametric inversion of strong lensing systems}, MNRAS 
{\bf 367} (2006) 1209, arXiv:\,astro-ph/0601124.
%
\bibitem{GA5}
S. A. Abel, A. Constantin, T. R. Harvey, A. Lukas, {\em Cosmic Inflation and Genetic Algorithms}, 
Fortsch. Phys. {\bf 71(1)} (2023) 2200161, 
arXiv:\,2208.13804[hep-th].
%
\bibitem{GA}
S. Nesseris, and J. Garcia-Bellido, {\em A new perspective on Dark Energy modeling via Genetic Algorithms}, JCAP {\bf 11} (2012) 033, arXiv:\,1205.0364 [astro-ph.CO].
%
%
\bibitem{gum}
B. Gumjudpai, T. Naskar, and M. Sami, and S. Tsujikawa, {\em Coupled dark energy: Towards a general description of the dynamics}, JCAP {\bf 06} (2005) 007, arXiv:\,hep-th/0502191.
%
\bibitem{mcp}
D. W. Pesce et. al., {\em The Megamaser Cosmology Project. XIII. Combined Hubble constant constraints}, ApJL {\bf 891}(1) (2020), arXiv:\,2001.09213 [astro-ph.CO].
%
\bibitem{low-matter}
S. Castello, M. Högås, and E. Mörtsell, {\em A cosmological underdensity does not solve the Hubble tension}, JCAP  {\bf 2022(07)} (2022) 003, 
arXiv:\,2110.04226 [astro-ph].
%
\bibitem{phantom}
L. Amendola, S. Tsujikawa, {\em Phantom crossing, equation-of-state singularities, and local gravity constraints in $f(R)$ models}, 
Phys. Lett. {\bf B} {\bf 660} (2008) 125-132, arXiv:\,0705.0396 [astro-ph].
%
\bibitem{phantom2}
M. Libanov, V. Rubakov, E. Papantonopoulos, M. Sami, and S. Tsujikawa, 
{\em UV stable, Lorentz-violating dark energy with transient phantom era}, 
JCAP {\bf 08} (2007) 010, arXiv:\,0704.1848 [hep-th].
%
\bibitem{growthdata}
B. Sagredo, S. Nesseris, and D. Sapone, {\em Internal Robustness of Growth Rate data}, Phys. Rev. {\bf D} {\bf 98}(8) 2018 (083543), arXiv:\,1806.10822 [astro-ph.CO].
%
\bibitem{sunny}
R. C. Nunes,  and S. Vagnozzi, {\em Arbitrating the S8 discrepancy with growth rate measurements from redshift-space distortions}, Mon. Not. Roy. Astron. Soc. {\bf 505}(4) (2021) 5427-5437, arXiv:\,2106.01208 [astro-ph.CO].
%
%
\bibitem{souza-neutrino}
D. H.F. de Souza, and R. Rosenfeld, {\em Can neutrino-assisted early dark energy models ameliorate the 
$H_0$ tension in a natural way?}, arXiv:\,2302.04644 [astro-ph.CO].
%
\bibitem{neutrino-tensions}
G. Garcia-Arroyo, J. L. Cervantes-Cota, U. Nucamendi, {\em Neutrino mass and kinetic gravity braiding degeneracies}, JCAP {\bf 08} (2022) 08, 009, arXiv:\,2205.05755 [astro-ph.CO].
%
\bibitem{chan-hTension}
M. H. Chan, {\em The cosmological ultra-low frequency radio background: a solution to the Hubble tension and the 21-cm excess trough}, Eur. Phys. J. {\bf C} {\bf 83} (2023) 6, 509.
%
\bibitem{daniel-s8tension}
P. D. Meerburg, {\em Alleviating the tension at low-$l$ through axion monodromy}, Phys. Rev. {\bf D} {\bf 90} (2014) 6, 063529, arXiv:\,1406.3243[astro-ph.CO].
%
\bibitem{clark-H0S8Tension}
S. J. Clark, K. Vattis, J. Fan, S. M. Koushiappas, 
{\em $H_0$ and $S_8$ tensions necessitate early and late time changes to $\Lambda$CDM}, Phys. Rev. {\bf D} {\bf 107} (2023) 8, 083527, arXiv:\,2110.09562 [astro-ph.CO].
%
\bibitem{AE-tension}
A. Alexandra, and G. Efstathiou, 
{\em A non-linear solution to the $S_8$ tension?}, 
Mon. Not. Roy. Astron. Soc. {\bf 516}(04) (2022) 5355, 
arXiv:\,2206.11794[astro-ph.CO].
%
\bibitem{KP-tens}
L. Kazantzidis and L. Perivolaropoulos, {\em Evolution of the $f\sigma_8$ tension with the Planck 15/$\Lambda$CDM determination and implications for modified gravity theories}, 
Phys. Rev. {\bf D} {\bf 97} (2018) 103503, arXiv:\,1803.01337[astro-ph.CO].
%
\bibitem{tsallis}
S. Basilakos, A. Lymperis, M. Petronikolou, E. N. Saridakis, 
{\em Alleviating both $H_0$ and $\sigma_8$ tensions in Tsallis cosmology}, arXiv:\,2308.01200[gr-qc].
%
\bibitem{amendola-GP}
A. Gómez-Valent, L. Amendola, {\em $H_0$ from cosmic chronometers and Type Ia supernovae, 
with Gaussian Processes and the novel Weighted Polynomial Regression method}, JCAP {\bf 04}
(2018) 051, arXiv:\,1802.01505 [astro-ph.CO].
%
\bibitem{pantheon}
D.M. Scolnic et al., The Complete Light-curve Sample of Spectroscopically Confirmed Type Ia Supernovae from Pan-STARRS1 and Cosmological Constraints from The Combined Pantheon Sample, arXiv:1710.00845.
%
\bibitem{ratra}
S. Cao, J. Ryan, B. Ratra, {\em Using Pantheon and DES supernova, baryon 
acoustic oscillation, and Hubble parameter data to constrain the Hubble 
constant, dark energy dynamics, and spatial curvature}, 
Mon. Not. Roy. Astron. Soc. {\bf 504}(1) (2021) 300--310, 
arXiv:\,2101.08817 [astro-ph.CO].
%
\bibitem{mifsud}
R. Briffa, C. Escamilla-Rivera, J. Levi Said, J. Mifsud, 
{\em Constraints on f(T) cosmology with Pantheon+}, 
Mon. Not. Roy. Astron. Soc. {\bf 522}(4) (2023) 6024--6034, 
arXiv:\,2303.13840 [gr-qc].
%
\bibitem{periv2}
G. Alestas, L. Kazantzidis, L. Perivolaropoulos, 
{\em $w-M$ phantom transition at $z_t$ \ensuremath{<}0.1 
as a resolution of the Hubble tension}, Phys. Rev. 
{\bf D} {\bf 103}(8) (2021) 083517, arXiv:\,2012.13932 [astro-ph.CO].
%
\bibitem{KPS-brs}
L. Kazantzidis, L. Perivolaropoulos and F. Skara, {\em Constraining 
power of cosmological observables: blind redshift spots and optimal 
ranges}, Phys. Rev. {\bf D 99} (2019) 063537, e-Print:\,1812.05356[astro-ph.CO].
%
\bibitem{bias}
S. Basilakos, J. B. Dent, S. Dutta,  L. Perivolaropoulos, and M. Plionis, {\em Testing General Relativity Using the Evolution of Linear Bias}, Phys. Rev. {\bf D} {\bf 85} (2012) 123501, arXiv:\,1205.1875 [astro-ph].
%
\bibitem{scol-SN}
D. M. Scolnic {\it et. al.}, {\em The Complete Light-curve Sample of Spectroscopically Confirmed SNe Ia from Pan-STARRS1 and Cosmological Constraints from the Combined Pantheon Sample}, Astrophys. J. {\bf 859} (2018) 2, 101, 
e-Print:\,1710.00845[astro-ph.CO].


\end{thebibliography}
\end{document}